\documentclass{article}
\input{epsf.tex}
\begin{document}
\begin{center}
{\Large \bf Two Regions of a Possible Drastic Change in the Structure of the Excited States of Any Nuclei}\\
\end{center}
\begin{center} V.A.KHITROV,  A.M.SUKHOVOJ,$^{*}$\footnote
{Corresponding author. E-Mail: suchovoj@nf.jinr.ru.} and LI CHOL
\end{center}
\begin{center} {\it
Laboratory of Neutron Physics, Joint Institute for Nuclear Research,
Dubna, Russia, 141980}
\end{center}
From a comparison of the absolute intensities of individual two-step cascades 
with known intensities of their primary and secondary transitions following 
thermal neutron capture the cascade and total population abilities of up to
$\sim 100$  levels  of each of the nuclei: $^{40}$K,
$^{60}$Co,$^{74}$Ge, $^{114}$Cd, $^{118}$Sn, $^{124}$Te, $^{137,138}$Ba, 
$^{150}$Sm,
$^{156,158}$Gd,
$^{165}$Dy, $^{168}$Er,
$^{175}$Yb, $^{181}$Hf, $^{183,184,185,187}$W, $^{196}$Pt and  $^{200}$Hg
have been determined.
These experimental data as well as the intensities of two-step cascades to 
the low-lying levels of these very nuclei can be restored within an accuracy
of experiment if only the level densities with a clearly expressed "step-like" 
structure are used and a considerable local increase of the radiative strength 
functions of secondary transitions to the levels situated close to the 
breakpoints on the energy dependence curve of level densities and their quite 
significant decrease to the low-lying levels of the nucleus are taken into 
account.
\section*{INTROUCTION}

The chief goal of experimental and theoretical investigations in low energy 
nuclear physics is the creation of a consistent model representation of the 
properties of nuclei in a specified interval of their excitation energy. A set 
of such models will provide an experiment-commensurable accuracy of the 
calculation of any practice-important parameters of the nucleus. 
To solve the problem, experiment must provide theory with a complex of 
experimental data that would reflect, in a quite explicit way, the most 
important properties of nuclear matter.
In practice, it is necessary that the density of the excited levels of nuclei in a specified interval of their quantum numbers over the entire investigated 
region of their excitation energy together with the emission probability of the 
products of the corresponding nuclear reaction should be determined. If the 
investigation is restricted to the binding energy of the nucleon in the nucleus, the main product of the nucleon capture reaction is gamma-quanta and the 
parameters determined are their radiative strength functions.

In nuclei with a sufficiently high density of excited levels ($\rho >10^{3}$
MeV$^{-1}$) it is practically impossible to single out an arbitrary gamma-transition and determine the situation of the levels it links.
This is why classical nuclear spectroscopy cannot solve the discussed problem.
However, neither the purposes of the development of nuclear theory nor of applied research require such detail information. It quite suffices to determine
experimentally the averaged parameters of the nucleus: the density of levels and the radiative strength functions of the gamma-transitions it emits.
Unfortunately, experimentalists have not been able to find a universal precise solution to the problem so far.

What seriously complicates the solution of the problem is that the intensity of 
the spectra of evaporative nucleons [1] and of primary gamma transitions emitted in nuclear reactions as excited levels with some arbitrary energy $E_{ex}$ 
discharge [2] does not depend, in any way, on the absolute values of the level 
density and emission probability of evaporative nucleons and gamma-quanta.

The situation has radically changed after  procedures for the determination of 
the intensities of two-step cascades as a function of the energy of their primary gamma-transition [3]:

 \begin{equation}
 I_{\gamma\gamma}=\sum_{\lambda ,f}\sum_{i}\frac{\Gamma_{\lambda i}}
 {\Gamma_{\lambda}}\frac{\Gamma_{if}}{\Gamma_i}=\sum_{\lambda ,f}
 \frac{\Gamma_{\lambda i}}{<\Gamma_{\lambda i}>
 m_{\lambda i}} n_{\lambda i}\frac{\Gamma_{if}}{<\Gamma_{if}> m_{if}},
\end{equation}
connecting the neutron resonance and specified low-lying levels of the studied 
nucleus and those for the extraction [4] from the data on the density of levels 
$\rho$ and radiative strength functions
\begin{equation}
k=\Gamma_{\lambda i}/(E_{\gamma}^3\times A^{2/3}\times D_{\lambda})
\end{equation}
of cascade gamma-transitions, were developed.
The value of  $I_{\gamma\gamma}$ is determined by the ratio of the partial widths of the primary $\Gamma_{\lambda i}$
and the secondary $\Gamma_{if}$ gamma transitions between the levels $\lambda$, $i$ and
$f$  to the corresponding total radiative widths and by the number of levels, 
$m$ and $n$, excited in the different intervals of their energies.
At the same time, their combination ensures (sooner qualitatively) the 
proportionality of the intensity to $\rho^{-1}$ and in this way, maximum 
sensitivity of the experiment to minimum level density values.

Since the difference between the spins of primary and final levels is not, as a 
rule, larger than 2, the value of $I_{\gamma\gamma}$ at thermal neutron capture 
is practically determined by only dipole transitions and level densities in 
unambiguously established intervals, $J^{\pi}$.

By now, the intensities of thermal neutron capture two-step cascades for all 
possible gamma-quanta energies have been measured for 51
nuclei from $^{28}$Al to $^{200}$Hg with coincidence spectrometers in Dubna, 
Riga, and \v{R}e\v{z}. For 40 of them, from the experimental spectra there was 
determined the portion of intensity that corresponds to the specified intervals 
of  primary gamma-transitions by applying the procedure [3].

Though the number of unknowns in Eq. (1) is always larger than the number of
the measured experimental values, the form of the dependence  $I_{\gamma\gamma}$
on its determining parameters allows the separation of finite (and rather small)
intervals of their values using which one can reproduce the experiment quite 
precisely. Naturally, the procedure [4] employs additionally known total 
radiative widths as well as known densities of low-lying levels and of neutron 
resonances.

A quite essential regularly observed difference between the observed and the 
calculated for 51 nuclei distributions of cascade intensity with the total 
energy $E_1+E_2=B_n-E_f$  (if the energies of their final level $E_f < 1$ MeV)
shows that the existing representations and models of cascade gamma-decay need to be seriously corrected.
We could not find any other possibility to make the accuracy of the model 
calculation of the cascade gamma-decay of any nucleus approach that of the 
present day experiment.

From analysis of all the data yielded by studies of two-step cascades 
(especially on the radiative strength functions of cascade transitions $k$ and 
on the density of the levels excited by them $\rho$ obtained in accordance with 
[4]) it follows that the structure of the wave functions of the excited levels 
is essentially different for their energy regions below and above $\sim 0.5B_n$. In the framework of today's theoretical representations of the dependence of 
the level density on the excitation energy of the nucleus the difference can be 
only explained (mainly qualitatively) as being due to the breakup of coupled 
nucleons.
Consequently, the level density and the probability of their excitation 
(discharge) differ significantly from those predicted on the basis of model 
representation of the nucleus as a purely fermion system (for example,[5,6]).

The importance of the conclusion comes from the fact that such model 
representations of the nucleus continue to be used today to analyze the 
experiment and to calculate gamma-spectra and neutron-nucleus interaction cross 
sections.
At the same time, the specific character of the data obtained in [4] together 
with analysis of the conditions of the corresponding experiment calls for going 
over not only to more realistic models of level density
$\rho$ and radiative strength functions (2)
(in a manner that would maximally reduce their dependence on the mass of the 
nucleus $A$)
(of the type [7,8] and [5,9], respectively),
but also to their more precise parameterization and further development.
The necessity of further mathematical developments is not only due to insufficient agreement between model representations and the experiment but is 
also due to demand for a more precise interpretation of processes occurring in the nucleus.

\section{ THE POSSIBILITY OF RELIABLE DETERMINATION OF
$\rho$ AND $k$ IN THE PRESENT-DAY EXPERIMENT}

Ordinary HPGe-detectors (relative effectiveness not higher than
30\%, no anti-Compton shielding) on a thermal neutron beam make it possible to 
obtain, in a period of
300 -- 700 hours of the experiment,  the cascade intensity distribution as a 
function of the energy of the primary transition in the cascade for the final 
levels with the excitation energy $E_f$ not higher than 1 MeV within an 
acceptable error and to determine the total intensity of two-step cascades to 
the upper lying levels with $E_f$ up to $\approx 2$ MeV.
This means to observe in the experiment, in the form of extremely simple and 
very convenient for analysis spectra of $\sim 50$  to $\sim 95 \%$ total 
intensity primary gamma-transitions at discharge of excited compound-states.

This allows obtaining of detail, precise and reliable information on the process of cascade gamma-decay on thermal neutron capture in any stable target nucleus. 
First of all, it is the density of the intermediate levels of the cascade and 
the radiative strength functions of cascade gamma-transitions.
Unfortunately, however, $\rho$ and $k$ obtained in accordance with [4] contain 
some unknown systematic error whose ordinary part is determined by
inaccuracies of neutron capture cross sections, measured gamma-quanta intensity 
in radiative thermal neutron capture spectra, particular experimental conditions of gamma-gamma coincidence registration, etc. 

In the present stage of technique development [4] the specific part of the 
systematic error is determined by possible existence of dependence of the 
strength function not only on the energy
$E_{\gamma}$ of a specified multipolarity quantum but also on the excitation 
energy $E_{ex}$ of the decaying level , not accounted for in [2] and [4], i.e., 
by the existence of the function  $k=F(E_{\gamma},E_{ex})$ in place of the 
assumption that $\Gamma_{i f}/\Gamma_{i}=F(E_{\gamma})$ made when Eq. (1) was derived.

The ordinary part of the determination error of $I_{\gamma\gamma}$
in the present-day experiment can be easily minimized to the level
not exceeding 5-20\%.
The specific part determines the difference, that is non-removable or difficult 
to remove today, between the extracted from experiment and real
$\rho$ and $k$.
 In principle, the difference can be completely removed by measuring the 
intensities of two-step cascades to final levels with energies not lower than 
3-5 MeV ( depending on the type of the investigated nucleus).
To do this, one can use multi-detector systems of HPGe-detectors whose number 
can even be limited.

The possibility of estimation of the effect of the decaying level energy $E_{ex}$ on the relative value
of the radiative strength functions of gamma-transitions of equal multipolarity 
and energy does exist at present.
For all the investigated nuclei there is obtained a considerable volume of 
information about the intensities

\begin{equation}
i_{\gamma\gamma}=i_1 \times i_2/\sum i_2,
\end{equation}
that are energy-resolved in the spectra as pairs of peaks of individual cascades.
Their parameters, including the most probable quanta ordering,
are reliably extracted from the experiment up to the cascade intermediate level 
excitation energy 3-5 MeV with the help of an original technique of analysis  
created in Dubna that employs a numerical algorithm   of resolution 
improvement
providing for a maximum possible resolution of all the obtained spectra  
$E_1+E_2=const$ without loss of effectiveness [10].

For $^{181}$Hf and even-odd isotopes of W, most complete intensity spectra of 
primary, $i_1$, and secondary, $i_2$, gamma-transitions emitted on thermal 
neutron capture were also measured up to $B_n$ in Riga and \v{R}e\v{z} [11].
The spectra of targets made of elements with a natural isotopic composition are given in [12]. The most fresh data on  $i_1$ and $i_2$ are in the file ``EGAF"
[13] in the amount that allows obtaining of quite acceptable
(though rather insufficient in volume) information on the discussed value for 
the nuclei $^{40}$K,
 $^{60}$Co,$^{74}$Ge, $^{114}$Cd, $^{118}$Sn, $^{124}$Te,  $^{137,138}$Ba,
$^{150}$Sm, $^{156,158}$Gd, $^{165}$Dy, $^{175}$Yb, $^{196}$Pt and  $^{200}$Hg.

Unfortunately, a limited volume of data on radiative thermal neutron capture 
spectra does not permit deriving of somewhat significant information for the 
rest 30 nuclei from the analysis discussed below. Among these nuclei, in the 
first place, are compound spherical and deformed odd-odd nuclei of middle mass. 

\section{METHOD OF DETERMINATION OF CASCADE POPULATION OF LEVELS}

From Eq.(3), taking advantage of the presently available data on $i_{\gamma\gamma}$, $i_1$  and $i_2$
the total population $P=\sum i_2$ of about $100$ levels  can be determined for 
the majority of the above enumerated nuclei to their excitation energy 3-4 MeV 
and higher. The difference between $P$ and the intensity $i_1$ of primary 
transitions to each of the levels is equal to the sum of their population by 2-, 3-, etc.-quantum cascades. 
It can be calculated in various ways if certain assumptions are made about the 
density of levels excited on thermal neutron capture and strength functions of 
cascade gamma-transitions.
To this end, there can be used, for example, the existing model representations 
of  densities and strength functions as well as possible hypotheses about them 
(including the values of $k$ and $\rho$ obtained in accordance with [4]).
Then the areas of maximum divergence of the experiment from the different 
calculation variants would show where and in what direction the model 
description of the cascade gamma-decay process should be modified.
Since at present, there is practically no possibility to determine 
experimentally the population of all, without exception, intermediate levels of 
two-step cascades even at their moderate excitation energies (due to the 
existing threshold of registration of the intensities $i_{\gamma\gamma}$, $i_1$ 
and $i_2$), it is reasonable to perform an experiment to calculation comparison
for $P-i_1$ summed over a small interval of excitation energies. Such sums 
should be looked at as a lower estimate for each of the intervals. We can make 
such a comparison for all of the enumerated compound nuclei. Simultaneously, the dependence of the intensity of two-step cascades on the energy of their primary 
transition $E_1$ is reproduced.

In the present-day experiment  [11-14], the error of $i_1$ and $i_2$ is limited 
by errors within which
the cross section of thermal neutron capture in the investigated isotope are known
[15]. In most cases it is not higher than 5-10\%. Accordingly, the population of any level from Eq.(3) has an accuracy that is only determined by the systematic 
error of the data [11-14], random error of particular  $i_1$ and $i_2$ and in 
addition, by errors of determination of gamma-quanta intensities in the 
HPGe-detector spectra due to partial overlapping of the peaks because of a 
limited resolution of the spectrometer. The portion of such cases can be reduced several times, by a maximum factor of 25\%, by careful approximation of neutron 
capture gamma-spectra in a practically monoisotopic target using the data on 
resolved two-step cascades.

To calculate $P-i_1$, the number of the dependence function variants of the 
strength functions and the density of levels on the gamma-quanta energy and the 
level excitation energy can be infinitely large. However, the general 
regularities 
of changes in the level population with their changing excitation 
energy can be determined using just three calculation variants:

a) The density of levels is predicted by any variant of the model of a 
noninteracting Fermi-gas,
the strength function of E1-transitions is specified by known extrapolations of 
the giant electric dipole resonance in the region below $B_n$,
and $k(M1)=const$ is specified by the normalization of $k(M1)/k(E1)$ to the 
experiment around $B_n$;

b) $\rho$ and $k$, that are obtained in accordance with [4] and reproduce 
exactly the intensity of two-step cascades as a function of energy of their 
primary transition, are used
(at present, only for the final levels in the cascades with $E_f <1$ MeV);

c) A set of level density and strength function values is chosen to reproduce 
exactly ($\chi^2/f <<1$) the values of $I_{\gamma\gamma}=F(E_1)$ (Fig. 1), the 
total radiative width $\Gamma_{\gamma}$ of the decaying compound state and the 
values of  $P-i_1$ at the same time.

The realization of the variant c) is possible in the iteration mode:
for $k$ obtained in accordance with [4] there is selected some dependence 
function that would change the secondary gamma-transition strength function 
values with respect to that of the strength function obtained in accordance with
[4] to enable the best reproduction of  $P-i_1$. To this end, it suffices to 
multiply the strength functions of the secondary gamma-transitions to the levels below some boundary excitation energy $U_2^{max}$ by the function $h$ containing several narrow peaks.
The dependence of their behavior on the excitation energy of the nucleus can be 
determined by analogy with the specific heat of ideal macrosystems in the second-order phase transition point as:

\begin{equation}
h=1+\alpha\times ({ln(|U_c-U_1|)-ln(|U_c-U|)})~~if~~U<U_c,
\end{equation}
\begin{equation}
h=1+\alpha\times ({ln(|U_c-U_2|)-ln(|U_c-U|)})~~if~~U>U_c,
\end{equation}
with some parameters $\alpha,~U_1,~U_2$ $U_c$. The condition $(U_c-U_1) \neq (U_2-U_c)$
ensures the necessary symmetry of the peaks and enables a somewhat more precise 
reproduction of the cascade level population at the tail-ends of the peaks in 
comparison with a Lorentz curve, for example.

In the best variant tested by us, the amplitude $\alpha$ must grow (linearly, for example)
as the excitation energy $U$ decreases from zero at $U=B_n$ to the maximally 
possible value shown in Figs. 2,3. The situations of the peaks, their amplitude 
and form are determined quite unambiguously by  $P-i_1$.
The population of any level whose number is $l$ is determined by the equation:

\begin{equation}
P_l=\sum_m P_m\times \Gamma_{m,l}/\Gamma_m,
\end{equation}
that depends on the population of all $m$ upper lying levels and on the
branching coefficient of their decay.
Although the data on the population depend on the two factors in the equation, 
the value of $P$ for the different low-lying levels is mainly determined by the 
relationship between the partial widths of the secondary transitions exciting 
them. Eq. (6) gives no other possibility to ensure an essential increase in the 
population of higher-lying levels.

The determined correcting functions are then included in the analysis [4] for 
the determination of $\rho$ and $k$ that exactly reproduce the cascade  intensities taking into account  an assumed difference between the energy  dependence of the strength functions of the primary and secondary transitions in  the cascade.
The values of  $I_{\gamma\gamma}$ are illustrated in Fig. 1 and the  re-determined level densities and strength functions are shown in Figs. 2,3.
 If necessary, the cycle is repeated once at most if the hypothesis of linearly  growing distortions in the value of $k(E1)$ and $k(M1)$ with increasing energy 
of the decaying levels is used and several times in case the hypothesis of
$\alpha=const$ is employed.
To minimize the number of the parameters to be selected, the correcting 
functions (4,5) are assumed to be similar for electric and magnetic 
gamma-transitions.
For the analyzed nuclei, the most general regularities in the behavior of the 
function $h$ retrace sufficiently well analogous dependence curves in [18].
In other words, there is observed a considerable increase of $k$ in the region 
of "step-like" structures and their considerable decrease for gamma-transitions 
to the low-lying levels.

The use of a large number of hypotheses is inevitable in the achieved stage of 
the problem being solved.
In this case, all the conclusions about the cascade gamma-decay process of the compound state should be considered as 
qualitative rather than 
quantitative.
So, the existence of a clearly expressed "step-like structure" of the level 
density and the related increase in $k(E1)+k(M1)$ (Fig. 2) can be considered as 
established with a high probability. However, the number and the form of such
"steps" may be only determined in further experiments. The same is true about 
the parameters of the correcting function $h$. Though the situation of the 
excitation energy region to which an essential increase in $k$ for 2nd-, 3rd-, 
etc.-transitions in the cascade corresponds causes no doubt (thanks to the 
number of tested variants), the particular parameters of the function $h$ should rather be considered as particularly preliminary and be used, in the first 
place, for the development of refining experiments.

\section{THE BEHAVIOR OF THE DEPENDENCE OF THE BEST $k$ AND $\rho$ ON
THE ENERGY OF THE DIPOLE GAMMA-TRANSITION AND THE EXCITATION ENERGY OF
THE NUCLEUS}

The realized method for the determination of $\rho$ and $k$ enables obtaining of their precise and reliable values employing practically no models.
Unfortunately, besides the analyzed sources of a possible systematic error, the $\rho$ and $k$ values may contain additional errors different for particular nuclei. For example, the absolute value of $k$ may have distortions due to some local deviation of the density of neutron resonances 
$\rho=D_{\lambda}^{-1}$ from its basic tendency or due to possible, though not accounted for in Eq. (1), structural effects.
This may be correlation between the partial radiative widths of cascade transitions and reduced neutron width of the resonance, that determines the basic part of the thermal neutron capture cross section.

The effects of those factors can be reduced by averaging radiative strength functions separately over even-even, even-odd and odd-odd compound nuclei. In the process of averaging there must be taken into account a rather strong difference 
between neutron binding energies in the investigated nuclei and very strongly differing dependence of the level density on the excitation energy of the nucleus.
In the variant suggested below, $B_n$ equals unity for each of the nuclei and the level density is taken in the form of its relationship with the simplest interpolating function
$const \times exp(\kappa E_{ex})$,
whose parameters are fully determined by the densities of neutron resonances and levels in the excitation energy region around 1-2 MeV.
Since $k$ presented by (2) depends weakly on the mass of the nucleus, the sum of the strength functions of dipole transitions is directly averaged over nuclei with equal parity of nucleons.
The averaging is performed for a set of a larger part of 40 nuclei for which
$\rho$ and $k$ are determined by the method [4] as well as for the nuclei whose population of individual levels is determined.
As it is seen from Fig.4, in the first and the second variant the energy dependence $k(E1)+k(M1) \approx const$ for the primary transitions with $E_1 < 0.3B_n$ independently of the nucleus type.
This confirms the principal validity of the basic representations of the model [3] for gamma-transitions from the compound states of high-lying levels. For odd-odd nuclei, however, strength functions are 2-3 times larger than similar data on Z-even nuclei.
Maximal possible values of $k(E1)+k(M1)$ are observed in the region
$E_1 \approx (0.7-0.8)B_n$ and they decrease as the primary transition energy further increases.

As it is seen from Fig. 5, the function $R=const \times \rho \times exp(-\kappa E_{ex})$ has maximums in the region of
$E_{ex} \sim 0.2$ and $\sim 0.8B_n$ and a minimum at about $0.5B_n$. From a comparison of two variants of the data for odd-odd nuclei (Fig. 5) it can be expected that their situations change as the mass of the nucleus changes (the population is only determined for $^{40}$K and $^{60}$Co) and most of $\rho$ and $k$ values are obtained for heavy deformed nuclei.
In the analysis [4], as it is accepted in such calculations, the entire excitation region of a particular nucleus is divided into ``continuous" and ``discontinuous " parts (with known scheme of decay). It also envisages the possibility of local variations of $k$ on the basis of experimental cascade intensity ``jumps" (Fig. 1). This leads to the appearance of ``breakups" in the functional dependence in Figs. 2 and 3.

A sufficiently general type of the dependence of the discussed parameters for nuclei having the different parity, $N$ or $Z$, makes it possible to conclude
that the extraction technique of $\rho$ and $k$ from the intensities of two-step cascades (employing data on the cascade population of levels in the nucleus)
allowing one to uncover the most general properties of the investigated parameters of the nucleus.

From the data in Fig. 5 it follows that in any type and/or mass nuclei (except for some nuclei) there exist at least two regions of excitation with a heightened density of levels. In spite of the common nature of how the regions demonstrate themselves, some additional modulation of radiative strength functions of the type shown in Figs. 4 and 5 can be assumed to exist at high excitation energies as well.
One can then assume, by analogy with Figs. 4 and 5, that more precise radiative strength functions of secondary transitions in the cascade should ensure
more exact reproduction of the data in Fig. 1.
This is especially important for well-deformed even-even nuclei for which the amount of the experimental data on the population of levels and the expected changes in strength functions is rather limited.

It should be noted that the conclusion made concerning a local increase of the radiative strength functions of the secondary gamma-transitions to the levels in the area of the "step-like" structure of the level density makes it possible to reproduce not only the basic peculiarities of the cascade population of levels below 3-4 MeV but also the dependence
$ I_{\gamma\gamma}=F(E_1)$
using practically similar dependence functions $\rho=\phi(E_{ex})$ and $k=\psi(E_1)$ for the different nuclei.
At the same time, their divergence from the existing models is much more expressed in comparison with [4] (Fig. 4).

The presented results should be looked at as a preliminary and, to some extent, qualitative description of the processes occurring in the nucleus. They cannot claim to be a complete and whole reproduction of the experimental picture of the cascade gamma-decay process because:

a) the hypotheses and model representations used in Eq. (1) may be inadequate to the experiment;

b) it is impossible to determine the number $N_c$ of the observed intermediate levels in the cascade with an error less than several tens of percent in the experiment performed for only one compound state (does not make it possible to exclude or estimate the degree of correlation between the neutron width of the compound state and partial radiative widths of secondary transitions);

c) it is impossible to estimate the total cascade population of the levels for which the value of $i_{\gamma\gamma}$ lies below the registration threshold and/or of the intermediate levels for which
$i_1$  and $i_2$ are unknown.

In spite of the above restrictions it is possible to conclude that the basic properties of the observed cascade decay process can be only reproduced within the framework of models that assume the existence of a considerable local increase of the radiative strength functions of gamma transitions to the levels lying in the interval with the width $\sim 1$ MeV in the vicinity of the effective excitation energy of a heavy deformed nucleus, 3-4 MeV.

\section{CONCLUSION}\

The attribute of the second-order phase transition is a sharp change of the internal properties of the investigated system as its energy changes.
While quite a sharp change of the level density (i.e., of the thermal capacity of the nucleus, in fact) was earlier established experimentally in [4] with a sufficiently high reliability, the results of the performed analysis point to a sharp change of the reduced probability of gamma-transitions (primary, at least)
in some, rather narrow, region of the levels of any nucleus excited by them.

The above reported results, that point to an essential increase in the radiative strength functions of secondary gamma-transitions for practically the same region of energies, can be considered as an additional independent proof of the existence of some region of excitation energies in the nucleus where a sharp change in the structure of the nucleus takes place. Presumably, it is a transition from domination of vibrational excitations to that of quasiparticle ones. Apparently, this can be interpreted as a phase transition from superfluid to ordinary state of such a specific system as nucleus. The effect is possibly associated with a breakup of the only pair of nucleons at excitation energies corresponding to a sharp decrease in the level density.

The whole set of the presently available data about the cascade gamma-decay of compound states excited on thermal neutron capture allows the conclusion that below the neutron binding energy a sharp change in the structure of the excited levels is observed at two excitation energies at least (Fig. 5).
The extrapolation of the conclusion to the region of high excitation energies results in the necessity of precise determination of level density using non-model independent procedures at high excitation energies as well.

Today, the data obtained as a result of the investigation of the nucleus in the discussed region of excitations should be rather considered as a preliminary indication to the possibility of the existence of such a transition. The quantitative information could be of use for planning of a more detail experiment to solve the discussed physical problem - a direct experimental investigation of the dynamics of the breakup of Cooper pairs in various finite nuclear systems dissimilar in the type of statistics and in their energy with respect to the Fermi surface.
\newpage
  {\bf REFERENCES} \\\\
1. M.I. Svirin, G.N. Smirenkin, {\em Yad. Fiz.\/} {\bf47} 84 (1988).\\
2. G.A. Bartholomew et al.,
{\em Advances in nuclear physics\/} {\bf 7} 229 (1973).\\
3. S. T. Boneva, V.A. Khitrov, A.M. Sukhovoj,
 {\it Nucl.  Phys.} {\bf A589}, 293 (1995).\\
4.  E.V. Vasilieva, A.M. Sukhovoj, V.A. Khitrov, {\it
Phys. At. Nucl.} {\bf 64(2)}, 153 (2001).\\
\hspace*{14pt} E.V. Vasilieva, A.M. Sukhovoj, V.A. Khitrov,
 {\it INDC(CCP)}, Vienna, {\bf 435}, 21  \\\hspace*{14pt}(2002)\\
\hspace*{14pt} http://arXiv.org/abs/nucl-ex/0110017\\
5. S.G. Kadmenskij, V.P. Markushev, W.I. Furman,
{\it Sov. J. Nucl. Phys.\/} {\bf 37}, 165  \\\hspace*{14pt}(1983).\\
6.  W. Dilg, W. Schantl,  H. Vonach,  M. Uhl, {\it
 Nucl. Phys.}, {\bf A217}, (1973), 269.\\
7. E.M.  Rastopchin, M.I. Sviron,  G.N. Smirenkin, {\it
Jad.Fiz.},  {\bf 52}, 1258 (1990).\\
8. A.V. Ignatiuk, Yu.V.  Sokolov, {\it 
Jad.Fiz.}, {\bf 19}, 1229 (1974).\\
9. V. A. Plujko,
Investigation of Interplay between Dissipation
 Mechanisms in \\\hspace*{14pt} Heated Fermi Systems by Means of
 Radiative Strength Functions,\\\hspace*{14pt}
 {\it Nucl. Phys.}, {\bf A649}, 209 (1999).\\
\hspace*{14pt} http://www-nds.iaea.or.at/ripl2/\\
10. A.M. Sukhovoj, V.A. Khitrov, {\it Instrum. Exp. Tech.}, {\bf 27} (1984) 1071\\
11. V. A. Bondarenko et al., Interplay of Quasiparticle and Phonon Excitations
 \\\hspace*{14pt}
in $^{181}$Hf Observed Through $(n,\gamma)$ and $(d(pol), p)$ Reactions,  \\\hspace*{14pt} {\it
Nucl. Phys.}, {\bf A709}  3 (2002).\\
\hspace*{14pt}
 P. Prokofjevs et al.,
Nuclear structure of $^{183}$W studied in $(n,\gamma)$,  $(n,n^{'}\gamma)$
\\\hspace*{14pt} and $(d,p)$ reactions
{\it Nucl. Phys.}, {\bf A614}, 183 (1997).\\
\hspace*{14pt}
 V. A. Bondarenko  et al.,
Nuclear levels in  $^{187}$W, {\it Nucl. Phys.}, {\bf A619}, 1  (1997).\\
12. M.A.  Lone, R.A. Leavitt, D.A. Harrison,{\it
Nuclear Data Tables}, {\bf 26(6)}, 511 (1981).\\
13. http://www-nds.iaea.org/pgaa/egaf.html\\  \hspace*{14pt}
G. L. Molnar et al.,
The new prompt gamma-ray catalogue for PGAA\\\hspace*{14pt}
{\it  App. Rad. Isot.},  {\bf 53}, 527 (2000).\\
14. V. A. Bondarenko, J. Honzatko, V. A. Khitrov,  A. M. Sukhovoj,
I. Tomandl, \\\hspace*{14pt}
Two-step cascades of the $^{185}$W compound nucleus gamma-decay,
\\\hspace*{14pt} // {\it  Fizika, B (Zagreb)}, {\bf 11}, 201 (2002).\\
15. {\it Neutron Cross Section}, vol. 1, part A, edited by S. F. Mughabhab,
 \\\hspace*{14pt} M. Divideenam, N. E. Holden, [Academic Press, New York,  1981].\\
16. P. Axel, Electric dipole ground transitions widths strength function
and 7 MeV \\\hspace*{14pt} photoninteractions,
{\it Phys. Rev.}, {\bf 126(2)}, 671 (1962). \\
17. J. M. Blatt, V. F. Weisskopf, {\it  Theoretical
Nuclear Physics}, [New York (1952)].\\
18. http://arXiv.org/abs/nucl-ex/0406030\\
\newpage

\begin{figure}
\vspace{-4cm}
\leavevmode
\epsfxsize=16.5cm

\epsfbox{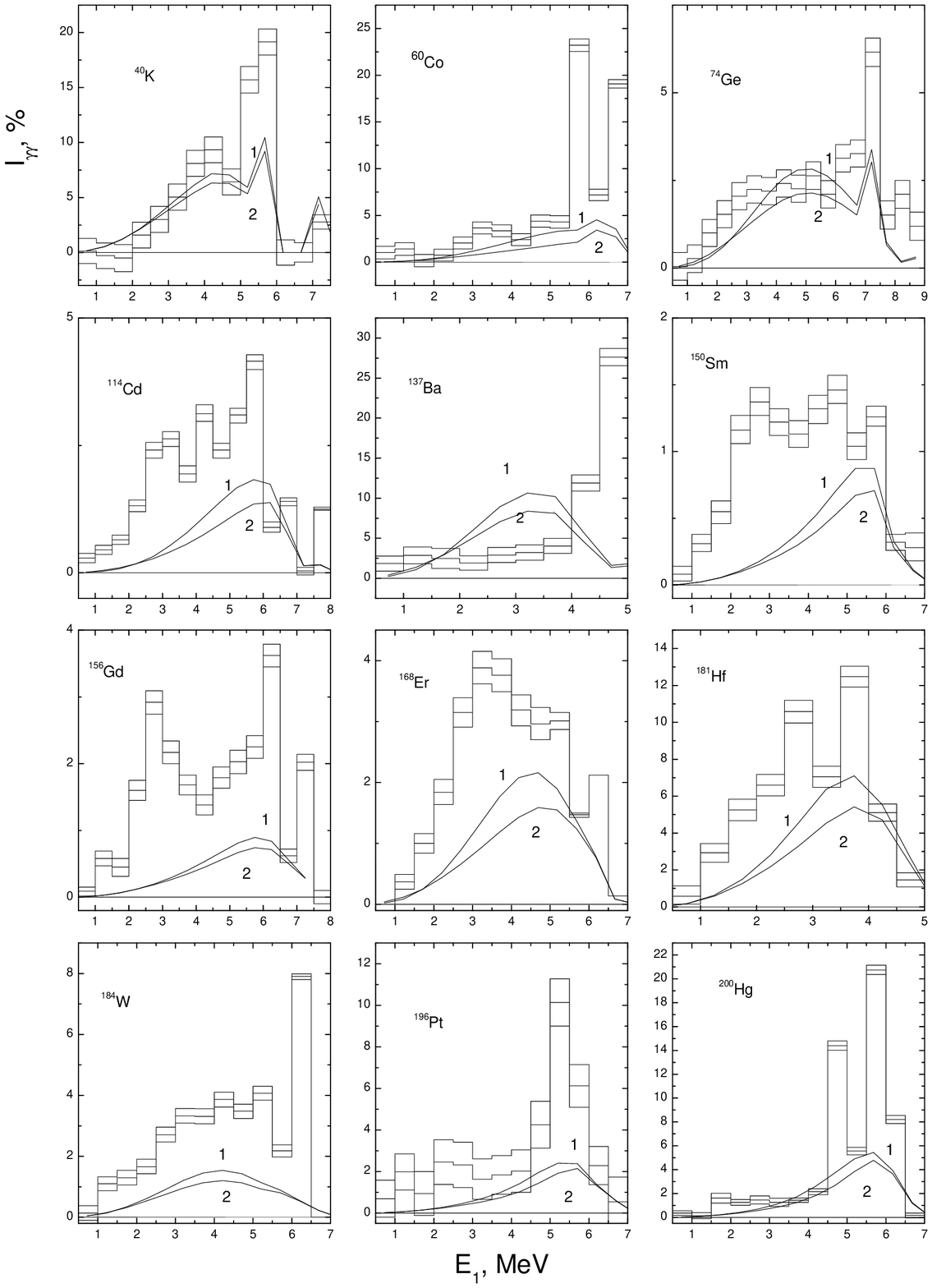}

\vspace{-2cm}
{\bf  Fig. 1.} {\it The examples of the intensity of two-step cascades as a function of their primary transition energy averaged over the intervals of the primary transitions energy $E_1$ with a width of 0.5 MeV for the nuclei $^{40}$K,
$^{60}$Co,$^{74}$Ge, $^{114}$Cd, $^{137}$Ba, $^{150}$Sm,$^{156}$Gd, $^{168}$Er,
$^{181}$Hf, $^{184}$W (renormalized to the data [13]), $^{196}$Pt and  $^{200}$Hg.
Curve 1 - calculated by Eq. (2) for the models  [4] and [16,17],
Curve 2 -- [5,6].}\\
\end{figure}
\newpage
\begin{figure}
\vspace{-4cm}
\leavevmode
\epsfxsize=16.5cm

\epsfbox{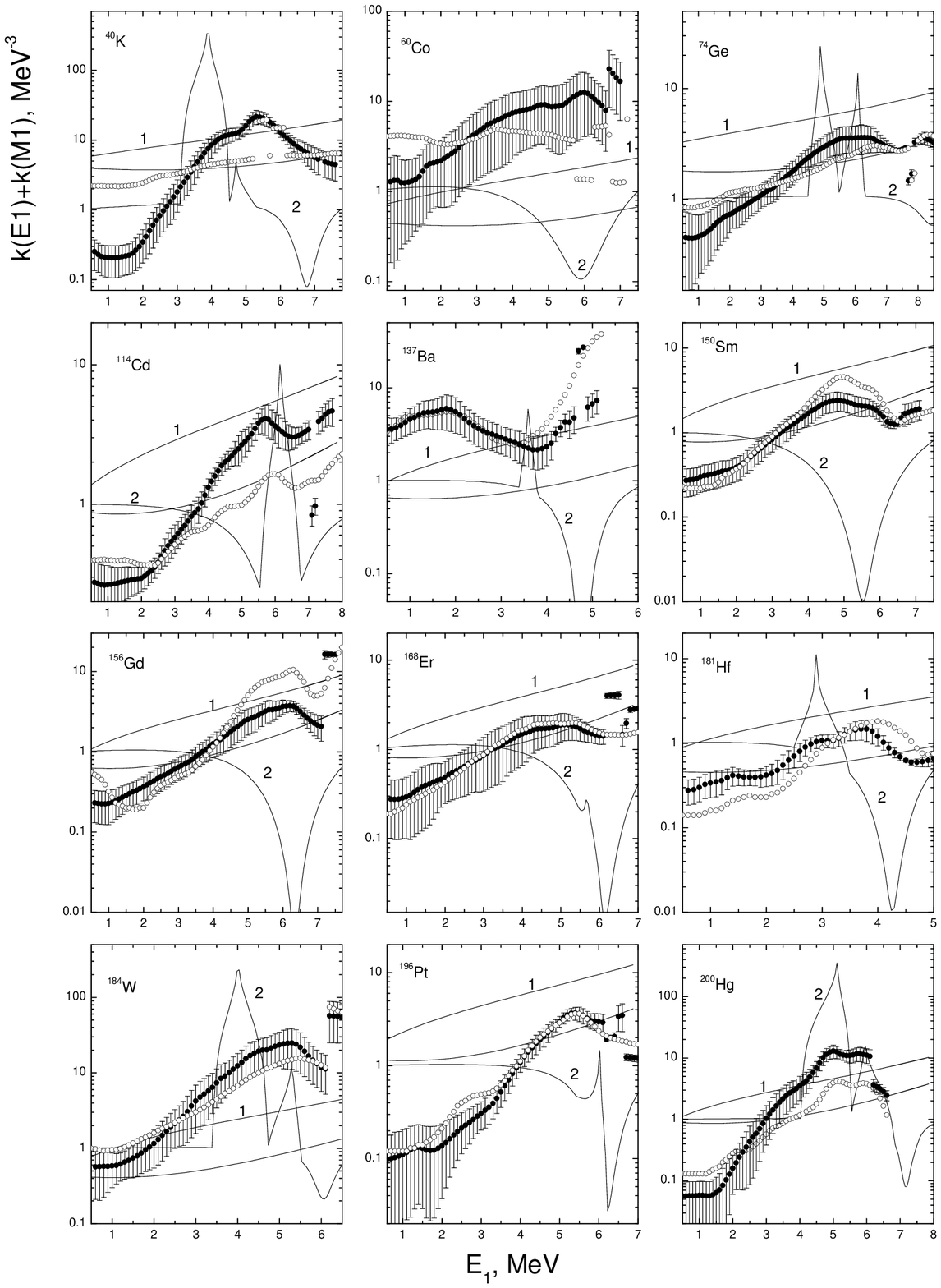}

\vspace{-2cm}
{\bf  Fig. 2.} { \it The points with errors - the sums of the radiative strength functions of the dipole electric and magnetic transitions in the cascade allowing precise reproduction of their intensities for the investigated difference of their values from the strength functions of the secondary transitions (multiplied by $10^9$). Open circles - similar values for the case of $h=1$.
Curve 1 (upper) - predicted by the model [16],  lower -- [5] under the assumption that $k(M1)=const$ [17].
Curve 2 - the maximum value of the function $h$ for the secondary gamma-transitions to the levels
 $E_i$.}\\
\end{figure}
\newpage
\begin{figure}
\vspace{-4cm}
\leavevmode
\epsfxsize=16.5cm

\epsfbox{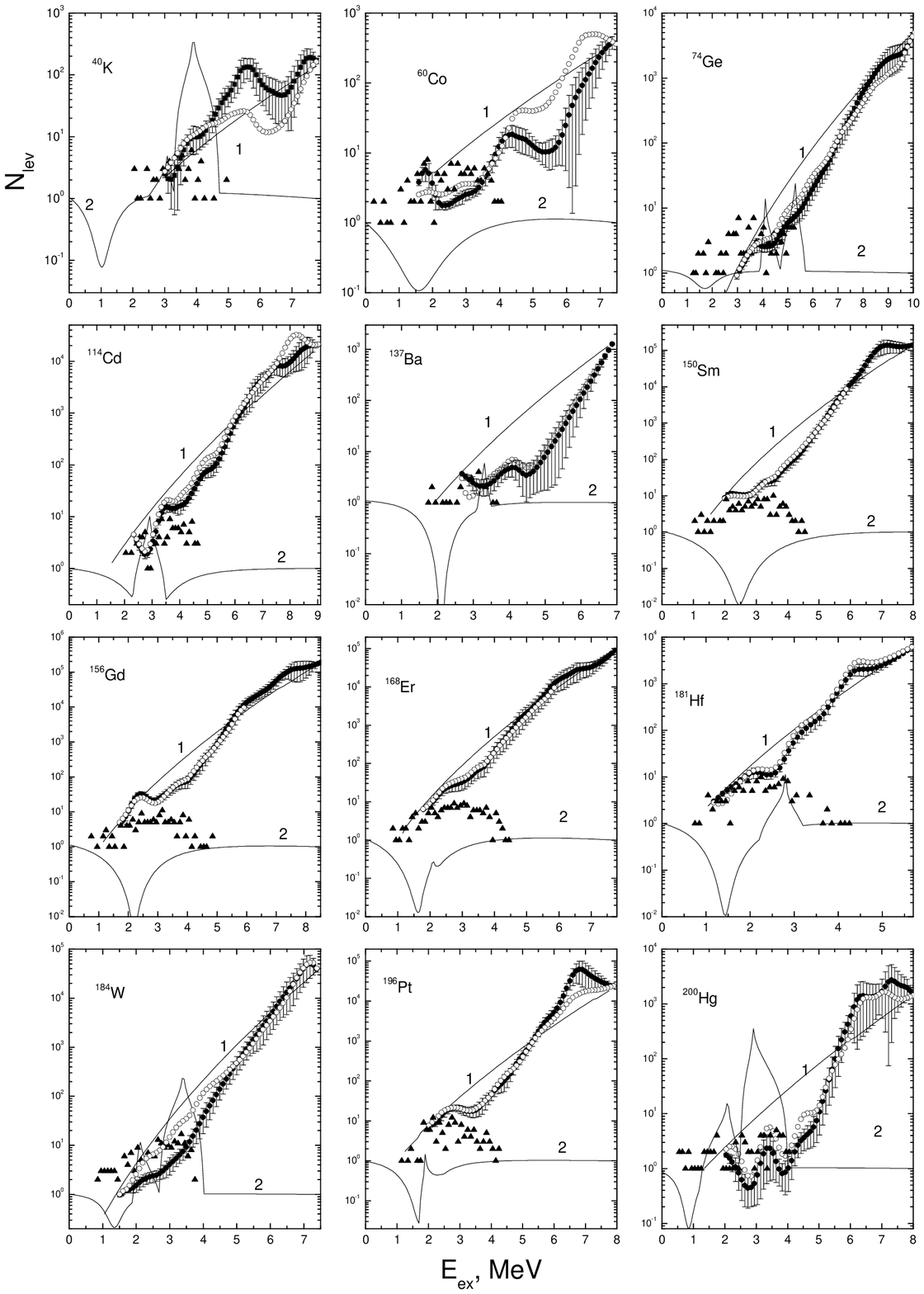}
\vspace{-2cm}

{\bf  Fig. 3.} {\it The points with errors - the total number of all intermediate levels in two-step cascades, that reproduces the whole set of the experimental data as a function of the intermediate level energy of the cascade.
Points - similar values for the case of $h=const$, the upper thin curve - predicted by the model [6].
The thick curve - the best value of the function $h$. Triangles - the observed number of intermediate levels in resolved cascades.}\\
\end{figure}\newpage
\begin{figure}
\vspace{-4cm}
\leavevmode
\epsfxsize=16.5cm

\epsfbox{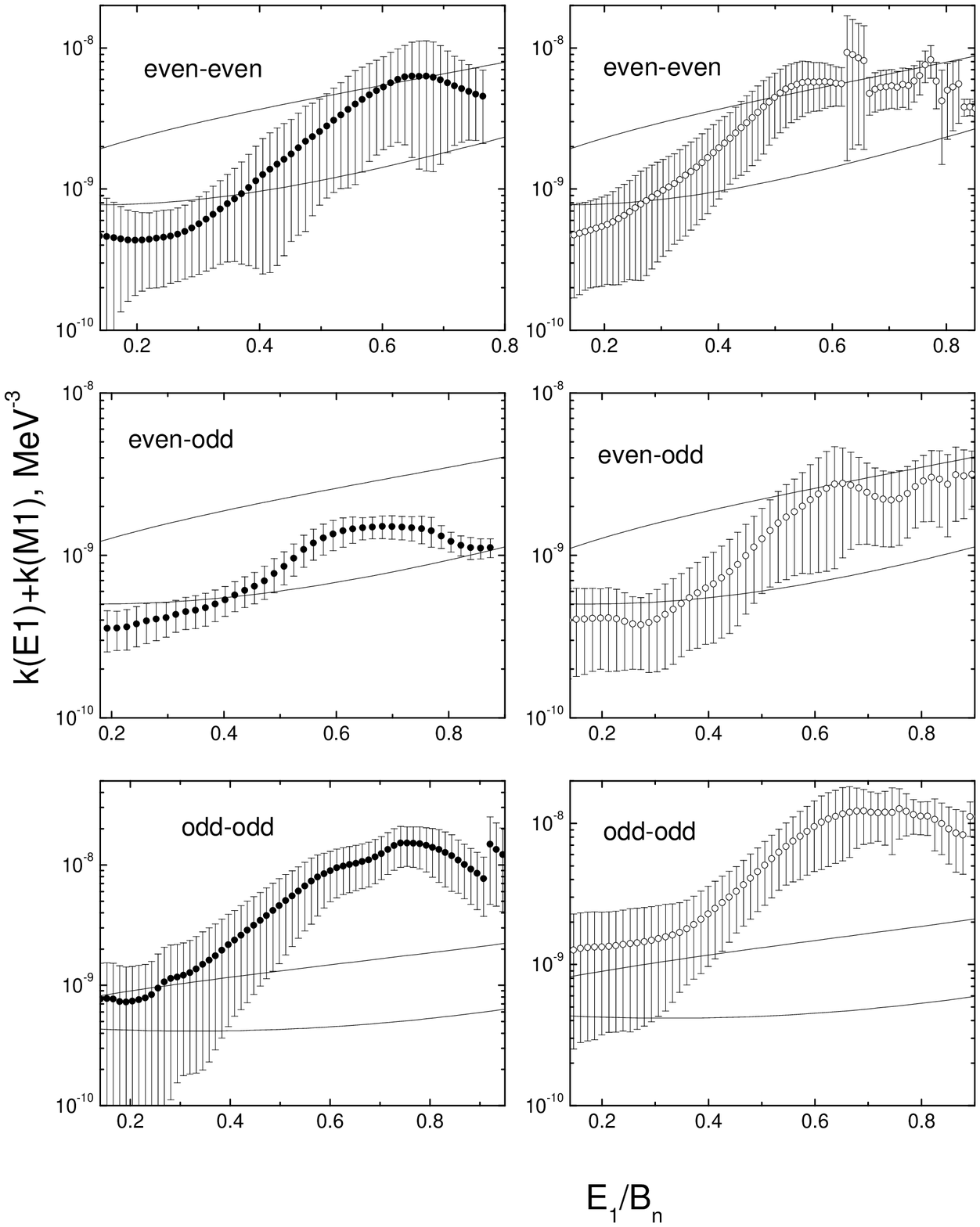}
\vspace{-2cm}

{\bf  Fig. 4.} {\it A comparison of the mean sums of the radiative strength functions of nuclei with the different parity of the number of neutrons and protons. Dark circles with errors - only nuclei for which the level population is determined. Open circles  - all the nuclei for which the analysis [4] is performed without accounting for the difference between the energy dependence of the strength functions of the primary and secondary transitions. 
Upper curve - predicted by the model [16], lower curve --[5] under the assumption that $k(M1)=const$}\\
\end{figure}
\newpage
\begin{figure}
\vspace{-4cm}
\leavevmode
\epsfxsize=16.5cm

\epsfbox{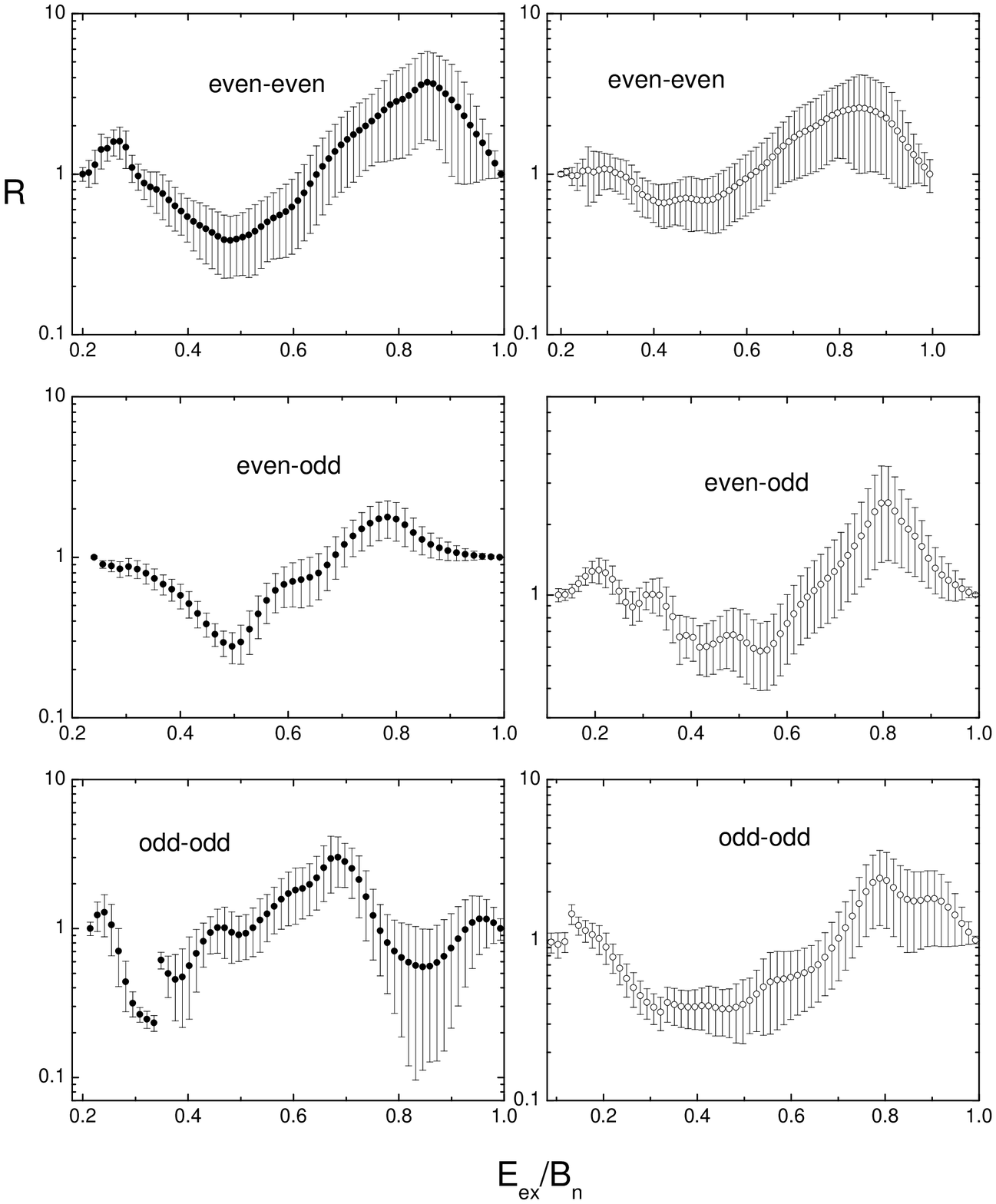}
\vspace{-2cm}

{\bf  Fig. 5.} { \it Mean relative variations of the level density. The notation is similar to that for Fig. 5.}
\end{figure}
\end{document}